\documentclass[noeprint,twocolumn,showpacs,amsmath,amssymb,sort&compress,floatfix,superscriptaddress,prl,aps]{revtex4-2}
\pdfoutput=1

\usepackage{graphicx}
\usepackage{dcolumn}
\usepackage{bm}
\usepackage[mathlines]{lineno}
\usepackage{mathtools}
\usepackage{amsmath}
\usepackage{float}
\usepackage[caption=false]{subfig}
\usepackage[english]{babel}
\usepackage[dvipsnames]{xcolor}
\usepackage{xparse}
\usepackage{float}
\usepackage{lipsum}
\usepackage{float}
\usepackage[normalem]{ulem}
\usepackage[hidelinks]{hyperref}

\DeclareMathOperator{\atantwo}{atan2}

\usepackage{etoolbox}

\makeatletter
\renewcommand{\fnum@figure}{FIG. \thefigure}
\makeatother

\newcommand{\add}[1]{{\color{OliveGreen}#1}}

\newcommand{\bl}[1]{{\color{blue}#1}}

\newcommand{\ue}{SUPA, School of Physics and Astronomy, University of Edinburgh, Peter Guthrie Tait Road, Edinburgh EH9 3FD, United Kingdom}

\begin{document}

\title{Anisotropic run-and-tumble-turn dynamics}

\author{Benjamin Loewe,}
\email[Corresponding author \\]{bloewe@ed.ac.uk}
\affiliation{\ue}
\author{Tyler N. Shendruk}
\affiliation{\ue}

%\affiliation{\ue}

\begin{abstract}
    Run-and-tumble processes successfully model several living systems. 
    While studies have typically focused on particles with isotropic tumbles, recent examples exhibit ``tumble-turns", in which particles undergo 90\textdegree tumbles and so possess explicitly anisotropic dynamics. We study the consequences of such tumble-turn anisotropicity at both short and long-time scales. We model run-and-tumble-turn particles as self-propelled particles subjected to an angular potential. Using agent-based simulations, we study the interplay of noise and potential on the particles' trajectories, demonstrating that the long-time effect is to alter the tumble-turn time, which governs the long-time dynamics. In particular, when normalized by this timescale, trajectories become independent of the underlying details of the potential. As such, we develop a simplified continuum theory, which quantitatively agrees with agent simulations. The hydrodynamic limit reveals that the transition to diffusive dynamics precedes the transition to isotropic dynamics as the hydrodynamic limit, while purely diffusive, can drive anisotropicity at intermediate times. 
\end{abstract}

\maketitle

\section{Introduction}

Since its inception, the field of active matter has relied on idealized models of self-propulsion to study the transport properties of living systems \cite{Vicsek1995, Romanczuk2012, Bechinger2016}. As persistent motion is often the main characteristic separating active particles from their passive counterparts, the unifying characteristic of these models is that particles are assumed to exert a propulsive force of constant magnitude but with a directionality that evolves in time through random processes \cite{Fodor2018, Bechinger2016}. As such, the specifics of their stochastic directional processes set these models apart from one another.

On the one hand, there are models with rotational diffusion. In these models, the orientation of the particle's self-propulsion follows a Wiener process and can be written as a Langevin equation. Examples include Active Brownian Particles (ABP) \cite{Redner2013, Fily2012, Marchetti2016} and Active Ornstein-Uhlenbeck Particles \cite{Bonilla2019,Farage2015,Maggi2015,Fodor2016}, which have inspired analogies with quantum systems \cite{Loewe2018,TeVrugt2023} and successfully captured dynamic processes in flocks \cite{Toner1998,Chardac2021a,Bain2019,Bricard2013a}, suspensions of motile Janus particles \cite{Walther2013,Ginot2018,Kurzthaler2018}, Quincke rollers \cite{Bricard2015,Liu2021,Chardac2021}, and cells within epithelial tissues \cite{Henkes2011,Giavazzi2018,Loewe2020,Bi2015,Perez-Gonzalez2018}.
On the other hand, there exist run-and-tumble models \cite{Schnitzer1993, Tailleur2008}. Run-and-tumble models mimic certain flagellar bacteria's characteristic motion (\emph{E. coli}) \cite{Berg,Soto2014,Saragosti2012}, particles randomly alternate between periods of constant self-propulsion (runs) and sudden sharp turning events (tumbles). The self-propulsion has a finite probability rate of instantaneously changing its direction, remaining constant otherwise. This time evolution makes writing a stochastic equation much harder than rotational diffusion. Recent attempts complement the Langevin equations with potentials that penalize deflections, thus retrieving effective run-and-tumble motion from a modified rotational diffusion \cite{Fier2018}. 
Although the distinction between ABP and run-and-tumble is inconsequential at long time scales \cite{Cates2013}, the different microscopic behavior can lead to striking emergent phenomena, such as phase separation \cite{Khatami2016}, when interactions between particles are significant.

Both APB and run-and-tumble processes rely on isotropy and only penalize angular deviations from a particle's current orientation. Nevertheless, active particles rarely lie in isolation, often interacting with external fields that explicitly break rotational symmetry and orient the particle within space. Examples include magnetic fields in abiotic \cite{Junot2022, Vach2017} and biological elements \cite{Waisbord2021,Vincenti2019,Hiscock2016}, re-orientations caused by shear flow \cite{Lee2021}, chemotaxis in cells \cite{Bhattacharjee2022}, bacteria \cite{Antani2021,Vladimirov2010} and synthetic systems \cite{Liebchen2018}, swimmers in nematic liquid crystals \cite{Goral2022} and cells crawling through patterned substrates \cite{Loy2021, Riching2015, Ray2017}.  Thus, such situations are expected to be anisotropic, depending on the particle's absolute orientation. 

\begin{figure}
    \includegraphics[width=\columnwidth]{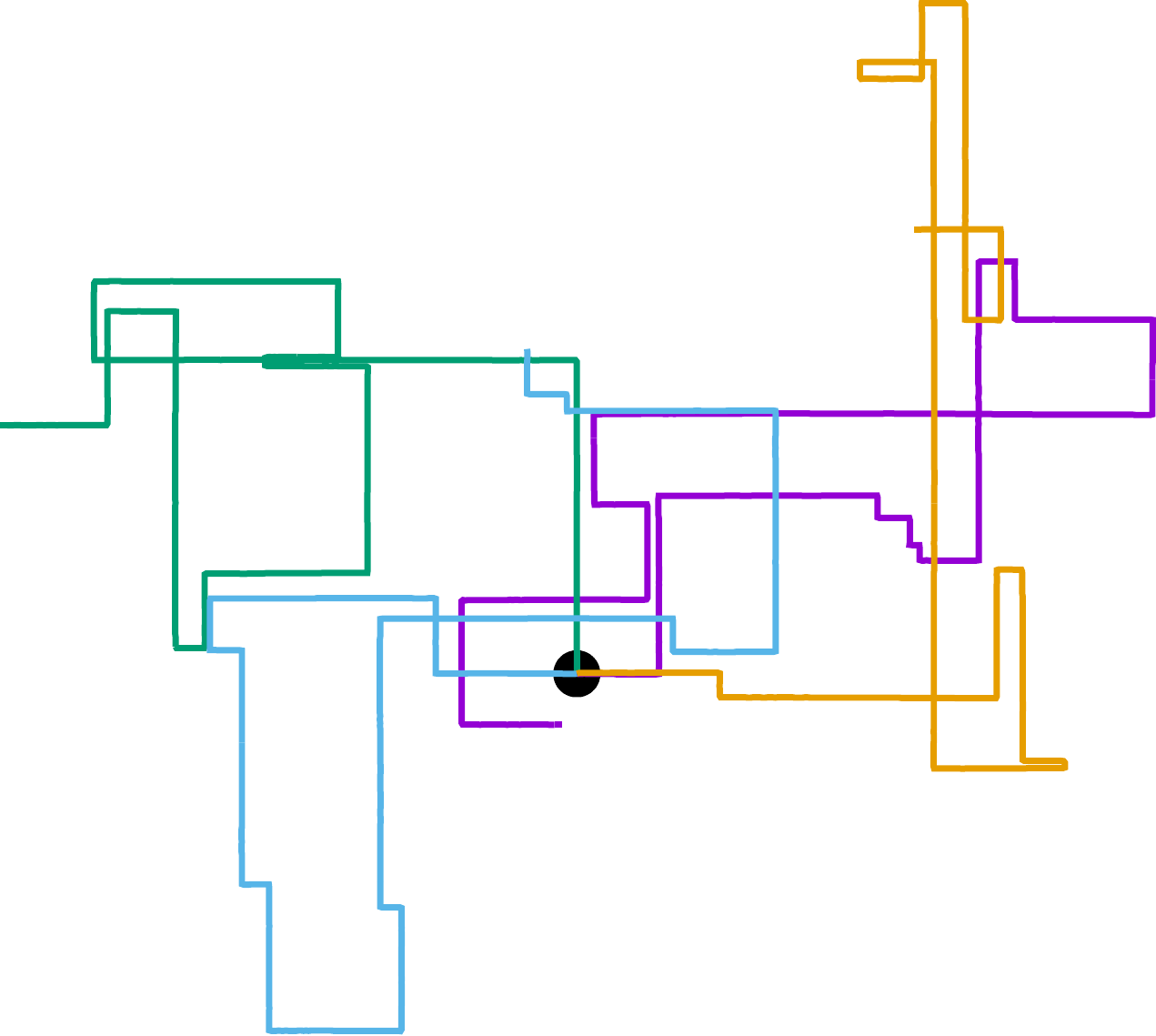}
    \caption{Particle trajectories ($\Gamma/D_r = 7$, $\delta=1.3$). The particles, which all have as a common starting point (the black circle), follow a run-and-tumble-turn process with sharp 90 degrees turns. 
    \label{fig:trajectories}}
\end{figure}

We recently introduced one such example: a Janus colloid with mixed anchoring conditions embedded in an active nematic \cite{Loewe2022}. This particle has planar anchoring on one side and homeotropic on the other. %with a continuous transition zone joining the two anchoring regimes. 
When immersed in an active nematic, the colloid effectively behaves as a self-propelled particle whose self-propulsion points either parallel or perpendicular to the local nematic director. Thus, this system represents an explicit realization of an effectively motile particle with externally imposed rotational symmetry breaking. The potential governing the colloid's orientation depends exclusively on the orientation itself and is entirely anisotropic. Activity-induced noise allows the colloid's orientation to hop from one potential well to another, leading to right-angled turns (\add{Fig.~}\ref{fig:trajectories}). The colloid resembles an effective anisotropic run-and-tumble\add{-turn} particle, for which tumbles must be sudden 90\textdegree left or right turns. While this occurs in two dimensions, recent work in active nematic droplets in three dimensions also exhibit pronounced right-angled turns \cite{Ruske2021}.

Inspired by the above example, we study the transport consequences of such a process, both analytically and computationally. To do so, we write a general Langevin equation for the particle's self-propulsion, which we use to simulate the run-and-tumble\add{-turn} dynamics. To obtain analytical results, we approximate this to a Poisson process and write equations for the particle's probability density. As the resulting model exhibits a hydrodynamical mode, we further coarse-grain the model to obtain a modified, explicitly anisotropic diffusion equation. By comparing these three descriptions, we find that, while highly anisotropic at short time scales with solutions having a markedly square geometry, all three descriptions become effectively isotropic at long time scales. While the Langevin and probabilistic description transition from self-propulsive at short time scales to diffusive at long time scales, the hydrodynamic theory is always diffusive, exhibiting that, although explicitly anisotropic, short wavelengths have a crucial role in shaping the propulsive behavior of the system. 

Moreover, the dynamics are characterized by a strong separation of scales: while the transition to diffusive behavior is marked by the time scale determined by the tumble-turn rate, at short time scales, the dynamics are governed by the relaxation time of the angular potential, resulting in a renormalization of the self-propulsion speed and a more or less noise short-time trajectory. When observed in the time scales and length scales associated with the particles turning, all trajectories are indistinguishable, regardless of the details of the angular potential. This strongly supports the use of a probabilistic continuum model. We perform a spectral decomposition of the probabilistic description and study the dynamic interplay of different wavelengths in shaping the propulsive-to-diffusive transition. Complementing this with a study of the hydrodynamic limit reveals that this transition precedes the transition to isotropic dynamics, with anisotropicity persisting in the diffusive regime.
These results expand the literature on anisotropic self-propelled processes, exhibiting anisotropicity of subtle origins that differ from anisotropic diffusion. Moreover, they shed light on how active systems with inherent anisotropic local dynamics can extrapolate this property over macroscopic time and length scales.

\section{Colloid Dynamics: Langevin Equation}

We consider self-propelled particles traveling at a constant speed $v_0$ with a direction specified by an orientation angle $\phi$ following a stochastic process. In contrast with the standard ABP, this process features an inherent anisotropy, incorporating a four-welled potential $V(\phi,\delta)$, favoring traveling along the orthogonal axes, $x$ and $y$. The dimensionless parameter $\delta$, which runs from $0$ to $\pi/2$, controls the width of the potential wells and, as such, quantifies how anisotropic this potential is. The equations governing these particles' position $\boldsymbol{r}(t)$ and orientation $\phi(t)$ are given by
\begin{align}
    \label{eq:Langevin_r}
    &d\boldsymbol{r} = v_0\, (\cos(\phi)\hat{\boldsymbol{e}}_x+\sin(\phi)\hat{\boldsymbol{e}}_y)\, dt,\\
    \label{eq:Langevin_Phi}
    &d\phi=\sqrt{2D_r}\,d\Omega_{t}-\Gamma \partial_\phi V(\phi,\delta) dt,
\end{align}
in which $\Gamma$ is a rotational mobility, $D_r$ a rotational diffusion coefficient and $\Omega$ denotes a Wiener process. 

The angular potential $V(\phi,\delta)$ is explicitly dimensionless, and $\Gamma$ has units of inverse time. Inspired by the anisotropic steering dynamics exhibited by the active nematic Janus colloids \cite{Loewe2022}, we consider the following rotational potential
\begin{align}
    \label{eq:V}
    &V(\phi,\delta) = \frac{f(2\phi-\pi/2,\delta)-f(\pi/2,\delta)}{f(\pi,\delta)-f(\pi/2,\delta)},\\
    &f(x,\delta) \equiv \text{Im}\left[\text{Li}_2\left(\frac{e^{2i(x+\delta)}}{\rho_0}\right)-\text{Li}_2\left(\frac{e^{2i(x-\delta)}}{\rho_0}\right)\right],
\end{align}
in which $\text{Li}_2$ denotes the di-logarithm and $\rho_0>1$ is a dimensionless parameter that smooths the potential, driving it away from divergent behavior. In \cite{Loewe2022}, this parameter corresponds to the distance between the colloid center and a companion topological -1/2 defect normalized by the colloid radius. Throughout, we fix $\rho_0=1.2$.
\begin{figure}
    \includegraphics[width=\columnwidth]{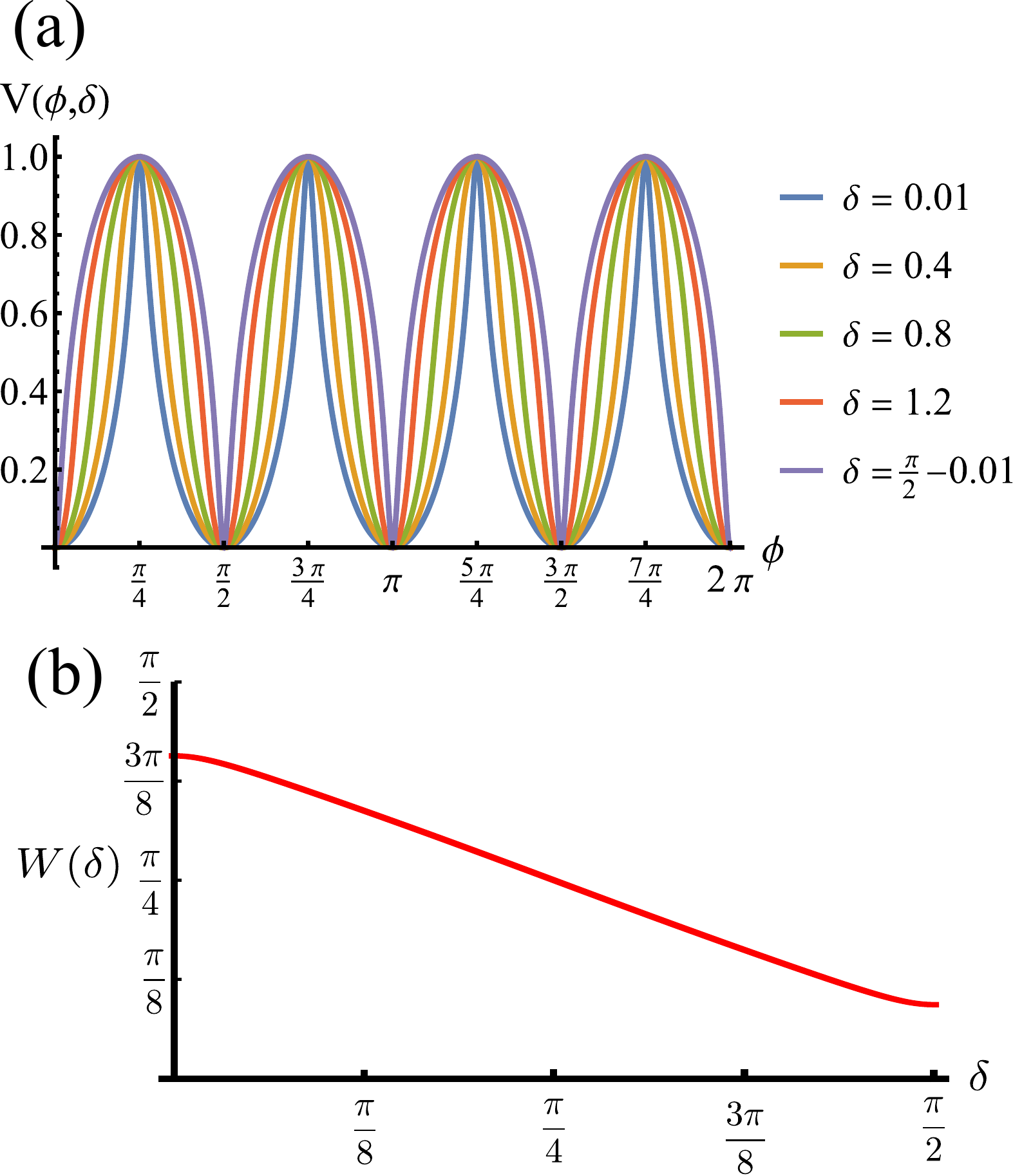}
    \caption{(a) Rotational potential as a function of $\phi$ for different values of $\delta$. The potential wells become narrower as $\delta$ increases.
    (b) Width of the potential wells $W$ as a function of $\delta$. The width is monotonically decreasing, behaving almost linearly. Because $\rho_0>1$ (see Eq.~(\ref{eq:V}), the width does not reach zero at any value of $\delta$, thus contributing to the numerical stability of our solutions for the Langevin equation. $V$, $W$ and $\delta$ are all dimensionless.
    \label{fig:potential}}
\end{figure}
The potential achieves its minimal value of $V=0$ at $\phi =0, \pi/2, \pi$, and $3\pi/2$ (Fig.~\ref{fig:potential}(a)). Similarly, its maxima occur at $\phi = \pi/4, 3 \pi/4, 5\pi/4$ and $7\pi/4$ with a value of $V =1$.  As such, by construction, the height of the potential barrier is $\Delta V = 1$, regardless of the value of $\delta$, with the angular mobility $\Gamma$ controlling the effective barrier height. However, the width of the potential wells $W$, defined as the angular difference between midpoints of the potential (i.e., $V=0.5$), decreases monotonically with $\delta$ (Fig.~\ref{fig:potential}(b)). 
The derivatives of the potential have compact closed expressions
%\begin{widetext}
%\begin{align}
%&\partial_\phi V(\phi,\delta) = \frac{4}{\Delta %f(\delta)}\text{Re}\left[\log\left(\frac{\rho_0+e^{i(4\phi-2\delta})}{\rho_0+e^{i(4\phi+2\delta)}}\right)\right],\\
%&\partial^2_{\phi}V(\phi,\delta) = \frac{1}{\Delta f(\delta)}\frac{32 r [2 r \cos(2\delta) + (1+r^2) \cos(4\phi]) \sin(2\delta)}{(1+r^2+2 r \cos(2 (\delta-2\phi)))(1+r^2+2 r \cos(2(\delta+2\phi)))},
%\end{align}
\begin{align}
    &\partial_\phi V(\phi,\delta) = \frac{4}{\Delta f(\delta)}\text{Re}\left[\log\left(\frac{\rho_0+e^{i(4\phi-2\delta})}{\rho_0+e^{i(4\phi+2\delta)}}\right)\right],\\
    &\partial^2_{\phi}V(\phi,\delta) = \frac{2 \rho_0 \cos(2\delta) + (1+\rho_0^2) \cos(4\phi)}{\Delta f(\delta)g(\delta-2\phi)g(\delta+2\phi)\csc(2\delta)},
\end{align}
in which we have defined $\Delta f(\delta) = f(\pi,\delta)-f(\pi/2,\delta)$, and $g(\theta) = (1+\rho_0^2+2\rho_0\cos(2\theta))/\sqrt{32\rho_0}$.

While Eq.~(\ref{eq:Langevin_r}) describes the self-propulsion of the colloid, Eq.~(\ref{eq:Langevin_Phi}) describes the particle's steering dynamics, with the particle's polar angle wiggling around one of $V$'s minima driven by noise, whose strength we quantify by $D_r$.
We begin by simply integrating Eqs.~(\ref{eq:Langevin_r}) and (\ref{eq:Langevin_Phi})  numerically for different values of $\delta$ using $\Gamma/D_r = 7.0$. For low values of $\delta$ (i.e., broader potential wells), the particles' orientations can explore a larger neighborhood around their minima. Thus we observe trajectories with a wider spread around the coordinate axes. However, these deviations are short-lived as the angular potential quickly pushes them back to the minimum. This leads to horizontal or vertical trajectories when viewed on long-length and -time scales. As $\delta$ is increased, the wells narrow, and the orientation becomes increasingly trapped near the potential minima of the potential. As a result, the particle follows well-collimated trajectories: straight lines with sharp right turns (Fig.~\ref{fig:trajectories}), thus giving the impression of the particle moving on a well-defined lattice.  
As such, the interplay between the angular potential and the angular noise results in two effects on the statistical behavior of the particle's dynamics: one at short timescales and the other at large time scales.

\subsection{Characteristic timescales}

\begin{figure}
    \includegraphics[width=\columnwidth]{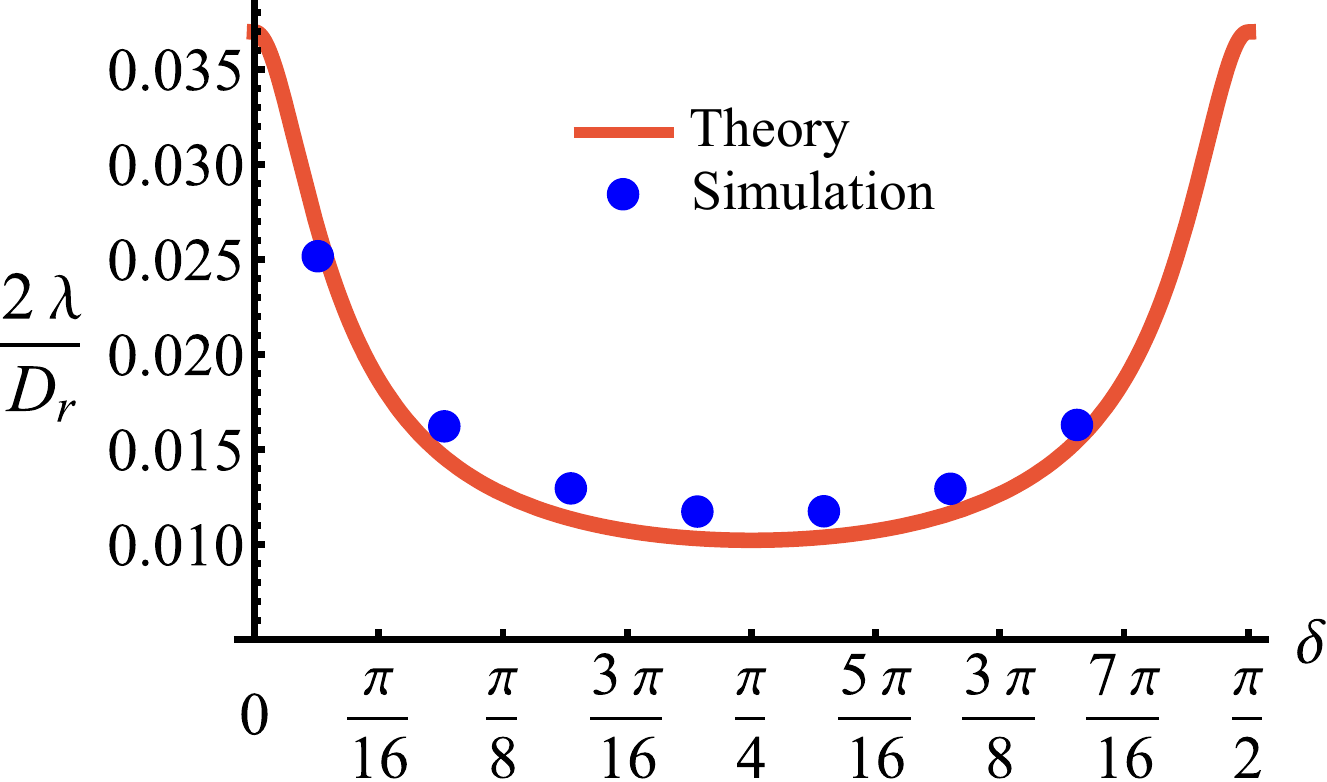}
    \caption{Total tumble-turn rate ($2\lambda$) as a function of the potential well's width $\delta$. The red curve corresponds to the theoretical approximation (Eq.~(\ref{eq:lambda})), whereas the blue dots are obtained by estimating the tumble-turn rate from the tangent-tangent correlation function (i.e., $\langle\cos[\phi(t)-\phi(t+\tau)]\rangle$). 
    \label{fig:lambda}}
\end{figure}

At short time scales, the orientation diffuses around the minima of the potential, leading to small deflections of the particle's velocity with respect to the system's axes. Depending on the width of the potential and the strength of the noise, this results in a less-collimated/broader distribution along the axes, with the particles taking longer to travel a fixed distance along the preferred axis.
Over longer time scales, however, the noise induces a finite probability for the polar angle $\phi$ to jump to a neighboring minimum. Any such jump amounts to a sharp right-angle turn, which amounts to a tumble-turn event in terms of being a sudden change to a previously persistent direction of motion. The timing between tumble-turn events corresponds to the mean time required to overcome the potential barrier $\tau$. As such, it will depend exponentially on the effective height of the barrier $\Gamma$, the strength of the noise $D_r$, and the curvature near the extrema. This is approximated using Krammer's formula
\begin{equation}
    \tau \approx \frac{2\pi}{\Gamma\sqrt{V''(\phi_\text{min},\delta)|V''(\phi_\text{max},\delta)|}} e^{\Gamma/D_r},
\end{equation}
in which $\phi_{\text{min}}$ and $\phi_{\text{max}}$ denote neighboring minima and maxima, respectively.
As such, the tumble-turn rate $\lambda = 1/\tau$ is found to be
\begin{equation}
    \label{eq:lambda}
    \lambda \approx \frac{16\Gamma}{\Delta f(\delta)\pi}\sqrt{\frac{\rho_0^2\sin^2(2 \delta)}{1+\rho_0^4-2 \rho_0^2 \cos(4\delta)}} e^{-\Gamma / D_r}.
\end{equation}
This tumble-turn rate only accounts for one neighboring minimum. Since each minimum is surrounded by two neighbors, the total tumble-turn rate is $2\lambda$. Per Eq.~(\ref{eq:lambda}), the only dependence of the turn rate on the shape of the well is through $\delta$. The tumble-turn rate increases substantially as $\delta$ approaches $\delta=0$ and $\pi/2$ (Fig.~\ref{fig:lambda}) because the curvature of $V(\phi,\delta)$ at its minima (maxima) increases substantially as $\delta\to\pi/2$ ($\delta\to0)$. 

In addition to the theoretical prediction \add{Eq.~}(\ref{eq:lambda}), Fig.~\ref{fig:lambda} also shows numerical estimations for $\lambda$, measured from the simulations as the characteristic time of the tangent-tangent correlation function, i.e., $\langle\cos[\phi(t)-\phi(t+\tau)]\rangle$. Both estimations agree with each other, with the theoretical curve having a slight underestimation for intermediate values of $\delta$.
%Since the derivation of Krammer's formula requires a Taylor expansion of the potential around its extrema, we do not expect Eq.~(\ref{eq:lambda}) to be valid at these increasingly singular values.
Nevertheless, the total tumble-turn rates provide a natural time scale $T = (2\lambda)^{-1}$ and length scale $\ell=v_0/(2\lambda)$.

\subsection{Long-time behavior}

\begin{figure}
    \includegraphics[width=0.9\columnwidth]{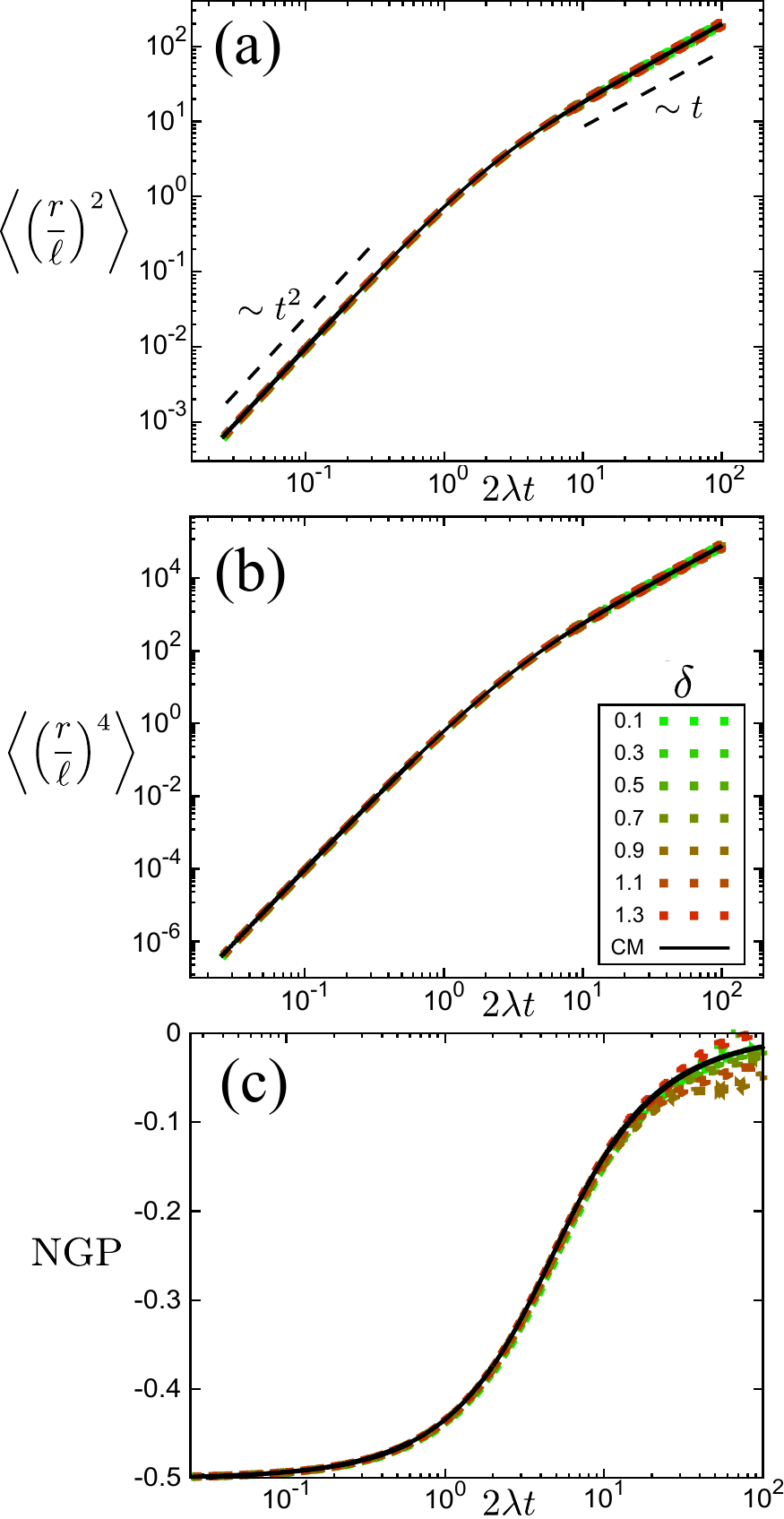}
    \caption{Normalized moments and non-Gaussian parameter $\text{NGP}$ from simulations of both the Langevin equation, Eqs.~(\ref{eq:Langevin_r}) and (\ref{eq:Langevin_Phi}), for various potential widths $\delta$ at $\Gamma/D_r = 7$ (dashed lines), and the continuum master equation, Eq.~(\ref{Eq:master_eq}) (solid lines, labeled CM). (a) Mean squared displacement ($\text{MSD})$, (b) fourth moment, and (c) the non-Gaussian parameter ($\text{NGP}$).
    \label{fig:moments}}
\end{figure}

The time scale $T$ defines what constitutes short and long times, as it controls in its entirety the shift from propulsive to diffusive behavior for the particle, as it can be observed by analyzing the moments of the particles' trajectories. 
Using the particle's position over time we compute the second moment ($\text{MSD}$), fourth moment and non-gaussian parameter ($\text{NGP}$, see Eq.~(\ref{eq:NGP}). It measures the extent to which the fourth moment deviates from the value of a Gaussian with the same second moment). Figure ~\ref{fig:moments} depicts these for various values of $\delta$ after rescaling them using $T$ and $\ell$ as our units of time and length, respectively. The resulting collapse of all curves into one for all these three measurements, Figs.~\ref{fig:moments}(a)-(c), clearly shows that the long-time behavior of the dynamics is controlled entirely by $T$, and therefore, by the tumble-turn rate $\lambda$. 

The $\text{MSD}$ (Fig.~\ref{fig:moments}(a)) shows a clear propulsive regime at $t\ll T$ and a transition to a propulsive regime for $t\gg T$. The propulsive dominated dynamics at small $t$ is also revealed by the negative sign of the $\text{NGP}$ (Fig.~\ref{fig:moments}(c)), which for a purely ballistic particle, has a value of $-0.5$.  As particles begin to turn and enter into a diffusive regime, their distribution becomes increasingly more isotropic, thus increasing the $\text{NGP}$. At very large $t$s, the dynamics are essentially diffusive, thus leading to a normal distribution, as highlighted by the $\text{NGP}$ approaching zero.

\subsection{Short-time propulsive behavior}

In addition to the turning dynamics that govern long-time behavior, the particle also experiments short-time wiggling around the minima of the angular potential. Even though this short time can broaden the particle's spatial distribution, it preserves propulsive dynamics. This is because a particle never travels back in the opposite direction, maintaining, on average, a constant direction of motion.

This hand-waving argument can be put into more formal grounds by studying the moments of the spatial probability distribution associated with Eqs.~(\ref{eq:Langevin_r}) and (\ref{eq:Langevin_Phi}). Since we are currently interested in studying the effects of noise near the minima and do not yet consider barrier jumping dynamics, a coarse approximation to the potential in Eq.~(\ref{eq:Langevin_Phi}) will suffice and, in what follows, a harmonic approximation is employed, i.e., $V(\phi,\delta) \approx 1/2 k\phi^2$, in which $k = \partial^2_\theta V(\phi=0,\delta)$.  With this approximation, Eq.~(\ref{eq:Langevin_Phi}) becomes the Langevin equation of an Ornstein–Uhlenbeck process, whose associated Fokker-Planck equation has a well-known solution \cite{Risken}
%\begin{equation}
%\label{eq:OU}
%\begin{split}
%P(\phi,t|\phi',t') = &\sqrt{\frac{k}{2\pi D_r (1-\exp(-2k (t-t')))}}\\ 
%&\exp\left[-\frac{k}{2 D_r}\frac{ (\phi-\phi'\exp(-k(t-t')))^2 }{ (1-\exp(-2k (t-t')))}\right].
%\end{split}
%\end{equation}
\begin{equation}
    \label{eq:OU}
    \begin{split}
        P(\phi,t|&\phi',t') = \sqrt{\frac{A(t-t')}{\pi}} \add{\times}\\ 
        &\exp\left[-A(t-t') (\phi-\phi'\exp\left[-k(t-t')\right])^2\right],
    \end{split}
\end{equation}
in which $A(t)\equiv k(D_r[1-\exp(-2 k t)])^{-1}$.
Since the general solution to Eq.~(\ref{eq:Langevin_r}) is
\begin{equation}
    \boldsymbol{r}(t) = v_0\int_0^t dt\,(\cos(\phi(t))\hat{\boldsymbol{e}}_x+\sin(\phi(t))\hat{\boldsymbol{e}}_y),
\end{equation}
the $\text{MSD}$ for a particle with its origin at its initial position is
\begin{equation}
    \langle r^2 \rangle = v_0^2\left\langle\int_0^t\int_0^{t} dt_2 dt_1 \cos(\phi(t_1)-\phi(t_2)) \right\rangle.
\end{equation}
Computing $\langle r^2 \rangle$ requires the probability of finding at time $t_2$ an orientation $\phi(t_2) = \phi_2$ knowing that at $t_1$ it had an orientation $\phi(t_1)=\phi_1$. 
%Defining
%\begin{align}
%    &A\equiv k/(2 D_r (1-\exp(-2 k t_1)),\\
%    &B\equiv k/(2 D_r (1-\exp(-2 k (t_2-t_1))),\\
%\end{align}
Using Eq.~(\ref{eq:OU}) with $A = A(t_1)$ and $B = A(t_2-t_1)$ and $\alpha \equiv  \exp(-k(t_2-t_1))$ this probability is
%\begin{widetext}
%\begin{equation}
%\begin{split}
%P(\phi_2,t_2;\phi_1,t1) &= \sqrt{\frac{k}{\pi \kappa (1-\exp(-2 k t_1))}} e^{- k \phi_1^2 / (1-\exp(-2 k t_1))}\\
%& \sqrt{\frac{k}{\pi \kappa (1-\exp(-2 k (t_2-t_1)))}} e^{- k (\phi_2-\phi_1 \exp(-k(t_2-t_1)))^2 / (1-\exp(-2 k (t_2-t_1)))},
%\end{split}
%\end{equation}
%\end{widetext}
% \begin{equation}
%     \begin{split}
%         P(\phi_2,t_2;\phi_1,t_1) = \frac{\sqrt{A B}}{\pi}
%          \exp(- A\phi_1^2-B[\phi_2-\phi_1 \alpha]^2).
%     \end{split}
% \end{equation}
\begin{equation}
    P(\phi_2,t_2;\phi_1,t_1) = \frac{\sqrt{A B}}{\pi} e^{- A\phi_1^2-B[\phi_2-\phi_1 \alpha]^2}.
\end{equation}
Thus, one must compute
\begin{equation}
    \begin{split}
        \langle r^2 \rangle = &2 v_0^2\int_{-\infty}^{\infty}\int_{-\infty}^{\infty}d\phi_1d\phi_2\int_0^t\int_{t_1}^{t} dt_1 dt_2\\
        &P(\phi_2,t_2;\phi_1,t1)\cos(\phi(t_1)-\phi(t_2)).
    \end{split}
\end{equation}
Performing the integrals over $\phi_1$ and $\phi_2$ leads to
\begin{equation}
    \langle r^2 \rangle = 2 v_0^2\int_0^t\int_{t_1}^{t} dt_1 dt_2 e^{-(A+[1-\alpha]^2 B)/4 AB},
\end{equation}
Changing the variables of integration $t_1$ and $t_2$ to $u=k(t_1+t_2)$ and $v=k(t_2-t_1)$, the integral becomes
\begin{equation}
    \langle r^2 \rangle = \frac{ v_0^2}{k^2} e^{-z}\int_0^{k t}dv \exp(ze^{-v})\int_{v}^{2 k t-v} du \exp(\beta z e^{-u}),
\end{equation}
in which $z=D_r/ k$, and $\beta = \cosh(v)-1$.  We now perform the inner integral and arrive at
\begin{equation}
    \label{eq:last_integral}
    \begin{split}
        \langle r^2 \rangle = \frac{ v_0^2}{k^2} e^{-z}\int_0^{k t} dv&e^{ze^{-v}}\left[\text{Ei}(\beta z e^{-v})
        -\text{Ei}(\beta z e^{-2kt+v})\right],
    \end{split}
\end{equation}
in which $\text{Ei}$ denotes the exponential integral function. 

To proceed forward, we recognize that for the case at hand ( i.e., for high enough barriers), fluctuations are expected to be smaller than $\pi/2$ (Fig.~\ref{fig:potential}(a)), such that the strength of the noise is much smaller than the force derived from the potential (i.e., $k$). Namely, the relaxation time to the minima of the potential $1/k$ is much shorter than the persistence time of the noise $1/D_r$. With this, we have that $z=D_r/k\ll 1$, and the arguments inside the exponential integral functions are much smaller than 1, allowing the expansion: $\text{Ei}(x) \approx \gamma_E+x(4+x)/4+\ln(x)$ (for $x\ll 1$), in which $\gamma_E$ denotes Euler's constant. Furthermore, after the expansion, the integrand can be well approximated by its linear expansion around $k\,t/2$. With the above approximations, the last integral in Eq.~(\ref{eq:last_integral}) can be evaluated. The result is a lengthy expression; however, as we are interested in the long-term consequences of the thermal movement of the orientation around the minima, it makes sense to consider the large $t$ limit, i.e., $t\rightarrow\infty$.  In this limit, the integral can be largely simplified to
\begin{equation}
     \langle r^2 \rangle = \frac{2 v_0^2}{k} e^{-z}t\left(kt+\frac{z}{2}+\frac{z^2}{16}\right).
\end{equation}
Since $z\ll 1$, this implies that, as $t\rightarrow\infty$
\begin{equation}
     \langle r^2 \rangle \approx v_0^2 e^{-z}t^2 = (v_0 e^{-D_r/(2k)}t)^2.
\end{equation}
In other words, even at long times the mean squared displacement grows propulsively with an effective velocity
\begin{equation}
    \label{eq:eff_v}
    v_\text{eff} = v_0\exp(-D_r/2k) \leq v_0.
\end{equation}

\begin{figure*}
    \includegraphics[width=\textwidth]{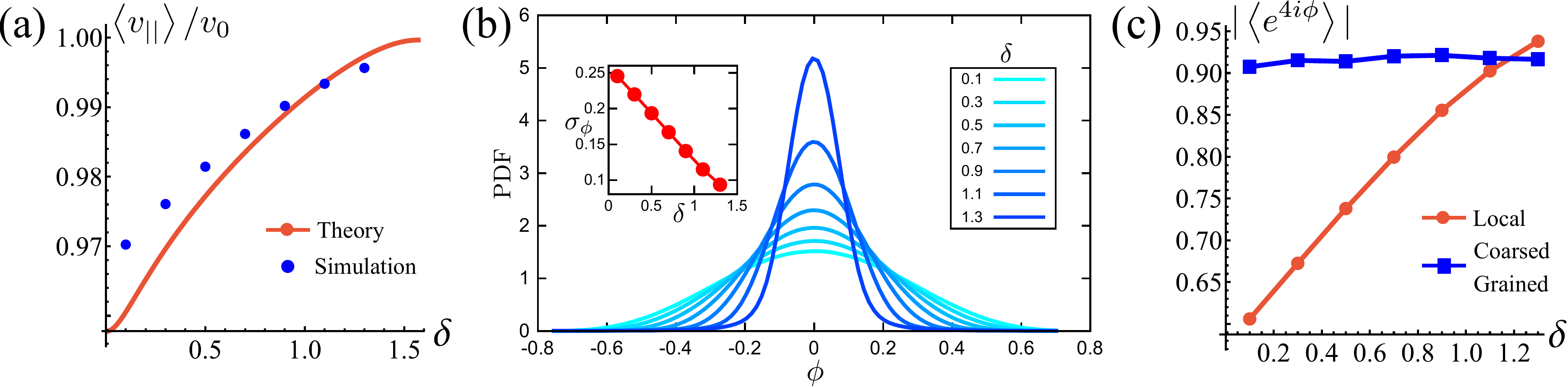}
    \caption{(a) Normalized effective velocity along the preferred axes $\langle v_{||}\rangle$. The red curve is the theoretical prediction from Eq.~(\ref{eq:eff_v}). Blue dots are measured directly from the simulation. (b) Distribution of instantaneous \add{polarization angle} $\phi$, translated to the interval $[-\phi/4,\phi/4]$, as function of potential width $\delta$. Inset: standard deviations as functions of $\delta$. (c) Measure of squarishness using instantaneous $\phi$ (red curve) and coarse-grained over $T$ $\phi$ (blue curve). 
    \label{fig:probs}}
\end{figure*}

Since $v_\text{eff} \leq v_0$, the random walk of the orientation along the potential minima does not induce a shift to diffusive behavior. The renormalization of the velocity occurs because the particle's velocity deviates further from its preferred direction as the noise increases, resulting in the particle translating more slowly along the axis. We can measure such an effect in our simulations by averaging over the component of the instantaneous velocities parallel to the instantaneously preferred traveling axis, $\langle v_{||}\rangle = v_\text{eff}$. Figure~\ref{fig:probs}(a) shows this estimate for different $\delta$'s and compares it with the theoretical prediction from Eq.~(\ref{eq:eff_v}). The estimations agree on the magnitude and in the monotonicity of $v_\text{eff}$ with $\delta$.

It is natural to ask how much the width of the potential affects this deviation from the preferred direction and how, in turn, this affects the squarishness of the particle's trajectories. To measure the directional deviations, we sample the distribution of instantaneous polarization angles $\phi$ (Fig.~\ref{fig:probs}(b)). This only shows angles between $-\pi/4$ and $\pi/4$, as akin to a first Brouillian zone, we have translated values of $\phi$ centered around other minima. It is clear that the shape of the distributions becomes increasingly sharper as $\delta$ increases, which is quantified by their linearly decreasing standard deviation, shown in the inset. This agrees with our notion of narrower potentials being more collimated.

It is striking, however, how broad the distribution appears to be at small $\delta$ (Fig.~\ref{fig:probs}(b)), as this would suggest that the trajectories are not very square-like. 
The answer to this apparent contradiction is that it is a matter of scales. Inspired by the bond orientational order parameter used to describe hexatic order \bl{\cite{Tanaka2019}}, we define the squarishness of a trajectory as $S \equiv |\langle \exp(4 i \phi)\rangle|$.
This directly measures the four-fold symmetry of our particle's trajectories: for isotropic trajectories $S\approx 0$, whereas for particles traveling mainly along lines intersecting at a $\pi/2$ angle $S\approx 1$.
The red line in Fig.~\ref{fig:probs}(c) depicts $S$, measured using the instantaneous values of $\phi$, as a function of $\delta$.  Besides showing that $S$ is strongly monotonically increasing with $\delta$, this figure shows that at small $\delta$, $S$ has intermediate values that are not consistent with squared trajectories but also incompatible with disordered trajectories. These values reflect that at small $\delta$ the particles have noisy trajectories that mostly travel along some preferred axes. As such, the trajectory will look very noisy and not very ordered when looking over small distances and short time scales. However, when viewed from afar, these fluctuations should cancel out, leaving behind only the directions of the preferred axes and, therefore, a high value of $S$. 

Indeed, if instead of using instantaneous values of $\phi$ to compute $S$ we use coarse-grained values obtained by computing running averages of $\phi$ for each trajectory over the tumble-turning time scale $T=(2\lambda)^{-1}$, we see that the coarse-grained $S$ is independent of $\delta$ and it has a high value, of the order of $0.9$, as depicted by the blue curve in Fig.~\ref{fig:probs}(c). Moreover, this implies that the trajectories are always squared when looked at from the length scale of the particle's persistent length, with the deviations only affecting the short-length-scale dynamics, which agrees with our qualitative intuition from Fig.~\ref{fig:trajectories}. 

In summary, there is a distinct separation of scales: {\it (i)} the random walk about the potential controls the short-time behavior, renormalizing the velocity, and {\it (ii)} at long-time scales, overcoming of the potential barrier and tumble-turning controls the dynamics, transitioning the behavior from propulsive to diffusive.
Moreover, when considered from the perspective of the persistent length, the dynamics and trajectories of these colloids become indistinguishable from one another, regardless of the width of the potential. Thus, a continuum theory that captures just the essentials of the run-and-tumble-turn dynamics, namely the right-angled turns, and disregards the details encompassed by the width of the potential is likely to capture most of the system's dynamics.

%$\theta_p = \theta'_p+\phi_c = \pi/2 + \pi/4-\Phi'/2 = \pi/2+\pi/4-(2\Phi-\pi/2)/2 = \pi/2-\Phi$ 

\section{Master Equation for a single particle}

The particles know the environment's inherent coordinate system, which drives particle dynamics to lie approximately parallel to one of the coordinate axes. This section presents a continuum master equation for these particles. For the sake of simplicity, we assume the limit of very narrow potential wells, i.e., $\delta \to \pi/2$ (Fig.~\ref{fig:potential}(a)), so the small variations in the particle's propulsion are negligible when its orientation deviates from the minima of the potential. As such, the particle can only move in four directions: rightwards, upwards, leftwards, and downwards.
As a consequence of assuming only sharp trajectories, the stochastic orientational dynamics described in the previous section result in the particle experiencing random sharp $\pi/2$ turn. Since, in the Langevin description, the jumps are restricted to neighboring potential wells, the particle cannot directly do a u-turn, reversing its direction. As such, the entirety of the hopping dynamics is encompassed by the tumble-turn rate $\lambda$ (Eq.~(\ref{eq:lambda})). 

With these simplifications in mind, the particles' self-propulsion dynamics reduce to a Poisson random walk, which can be described as probability densities.
As the colloid can only move in four directions, there are four probability densities: $P_{\rightarrow}(\bm{r},t)$, $P_{\uparrow}(\bm{r},t)$, $P_{\leftarrow}(\bm{r},t)$, $P_{\downarrow}(\bm{r},t)$, which denote the probability of having a colloid in a neighborhood $d^2r$ of position $\bm{r}$ at time $t$ traveling right, up, left or down, respectively. Since the particle always travels at the same speed $v_0$ and turns with probability rate $\lambda$
 \begin{figure*}[t!]
    \includegraphics[width=\textwidth]{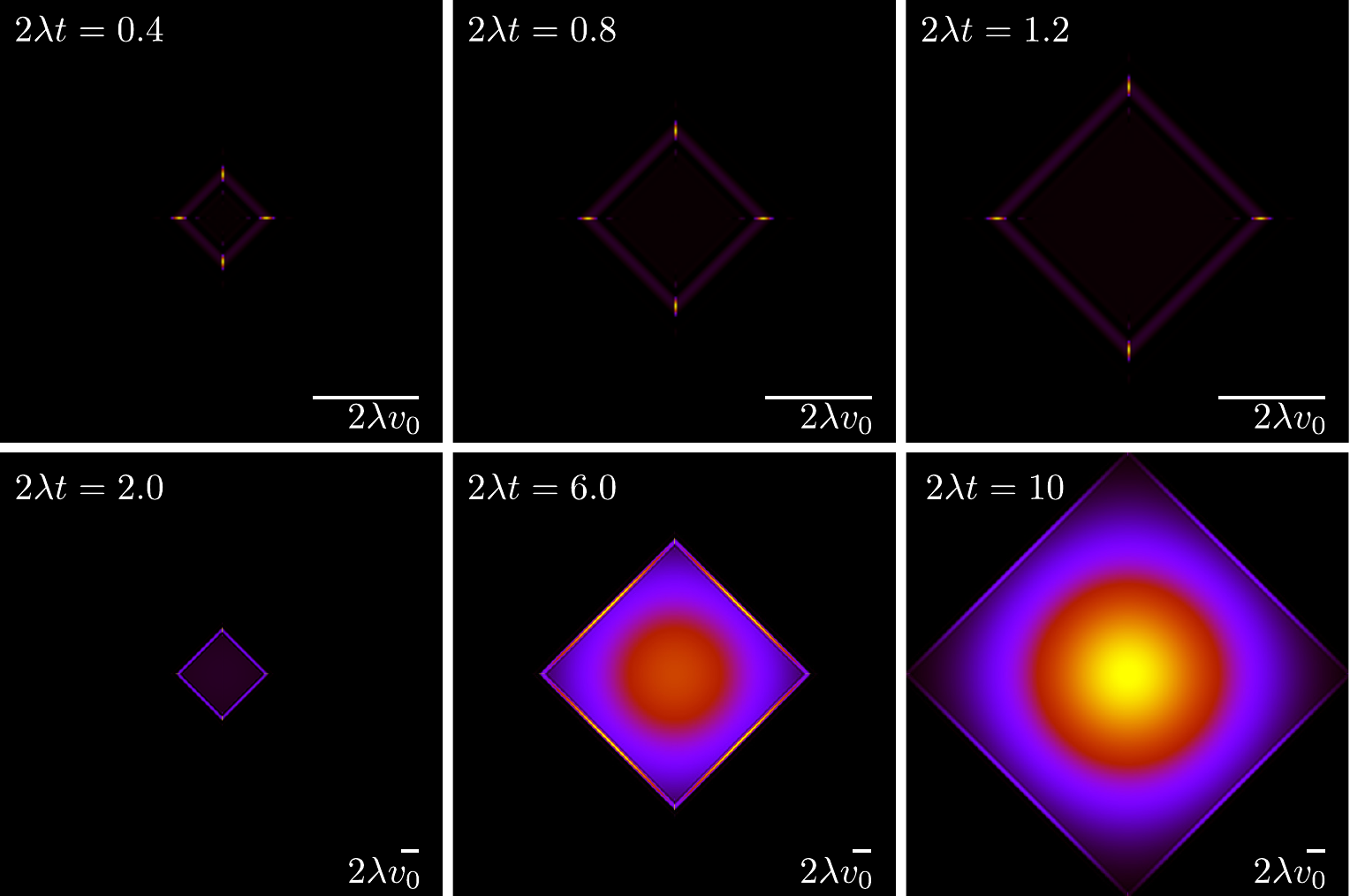}
    \caption{Time evolution of the total probability $P = P_{\rightarrow}+P_{\uparrow}+P_{\leftarrow}+P_{\downarrow}$ Eq.~(\ref{Eq:master_eq})), of an initialy well-localized distribution. The upper row (short times) has a different spatial scale than the lower row (long times).
    \label{fig:dynamical_dist}}
\end{figure*}
\begin{equation}
    \begin{split}
        P_{\rightarrow}(x,y,t) =  &    \,P_{\uparrow}(x,y-v_0\,dt,t-dt) \lambda \,dt \\
        & + P_{\downarrow}(x, y-v_0\, dt, t-dt) \lambda \,dt \\
        &+ (1-2\lambda\, dt) P_\rightarrow(x-v_0\,dt, y, t-dt).
    \end{split}
\end{equation}
This simply states that the probability of finding a particle at position $\bm{r}$ at time $t$ traveling to the right is the sum of the probabilities of having the same particle at an earlier time $t-dt$, with $\lambda\, dt \ll 1$, traveling either up or down and turning, or traveling to the right and not turning. There are similar expressions for the remaining three probability densities, which become exact in the limit $dt\rightarrow 0$. Expanding these expressions in $dt$ and keeping only terms up to first order produces the dynamical equation for the probability spinor $\bm{P} = (P_\rightarrow, P_\leftarrow, P_\uparrow, P_\downarrow$). Taking $T=1/(2\lambda)$ and $\ell=v_0/(2\lambda)$ as the unit of time and length, respectively, this dynamical equation takes the following dimensionless form
\begin{equation}
    \partial_t\bm{P} = -\boldsymbol{H}_D\boldsymbol{P},
    \label{Eq:master_eq}
\end{equation}
in which the dynamical matrix is
\begin{equation}
    \boldsymbol{H}_D = \begin{pmatrix}
    \sigma_z\,\partial_x+1 & -(1+\sigma_x)/2 \\
    -(1+\sigma_x)/2 & \sigma_z\,\partial_y+1
    \end{pmatrix}	
    \label{Eq:dyn_matrix}
\end{equation}
 for the Pauli matrices $\sigma_z$ and $\sigma_x$.
 
The evolution for the total probability $P = P_{\rightarrow}+P_{\uparrow}+P_{\leftarrow}+P_{\downarrow}$ under Eq.~(\ref{Eq:master_eq}), with a localized initial distribution without any preferred axis, is depicted in Fig.~\ref{fig:dynamical_dist}. At short times (upper row of Fig.~\ref{fig:dynamical_dist}), when propulsive dynamics dominate, the distribution has a marked 4-fold symmetry, with 4 probability peaks traveling along the axes with speed $v_0$.  As time reaches the characteristic time $T$, most particles have turned on\add{c}e, forming diagonal fronts that join the probability peaks. At larger times (lower row of Fig.~\ref{fig:dynamical_dist}), particles have had enough time to turn at least twice and do u-turns, thus returning to their initial position, incrementing the relative value of the probability distribution at the origin. At the same time, most particles have turned at least once, moving the maxima of the distribution out of the traveling peaks into the center of the diagonals.  Finally, at long times, the dynamics are diffusive, leading to a normal distribution at an increasingly larger distance near the origin. At length scales comparable with $v_0 t$, the probability distribution still has a 4-fold symmetry due to the diagonal front of particles that have turned at most once. 

 \begin{figure*}
    \includegraphics[width=\textwidth]{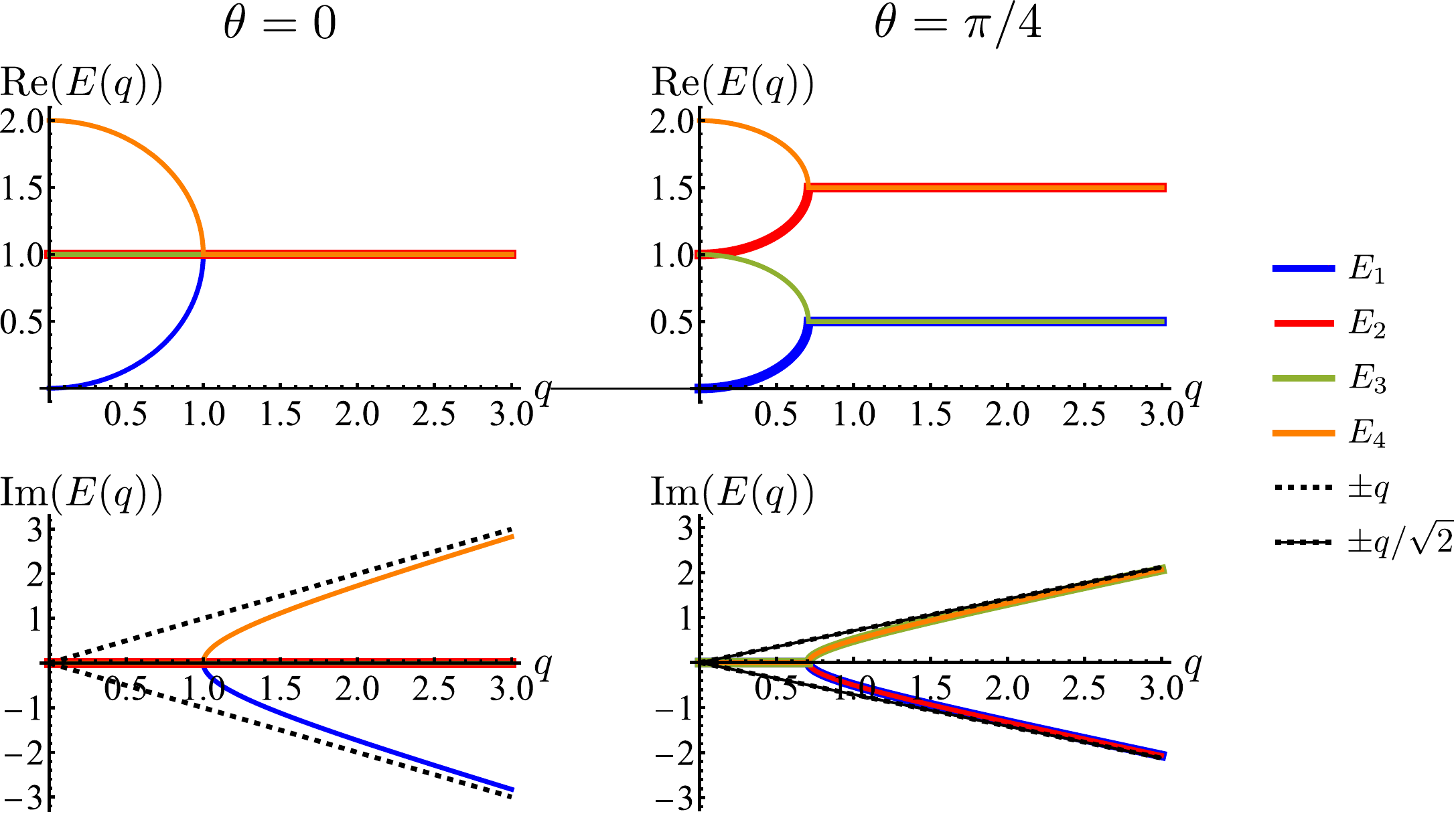}
    \caption{Real and imaginary parts of the eigenvalues of the dynamical matrix, Eq.~(\ref{Eq:eigenvalues}), as functions of the radial wave-number $q$ and its polar angle $\theta$
    \label{fig:eigenvalues}}
\end{figure*}
\begin{figure*}
    \includegraphics[width=0.9\textwidth]{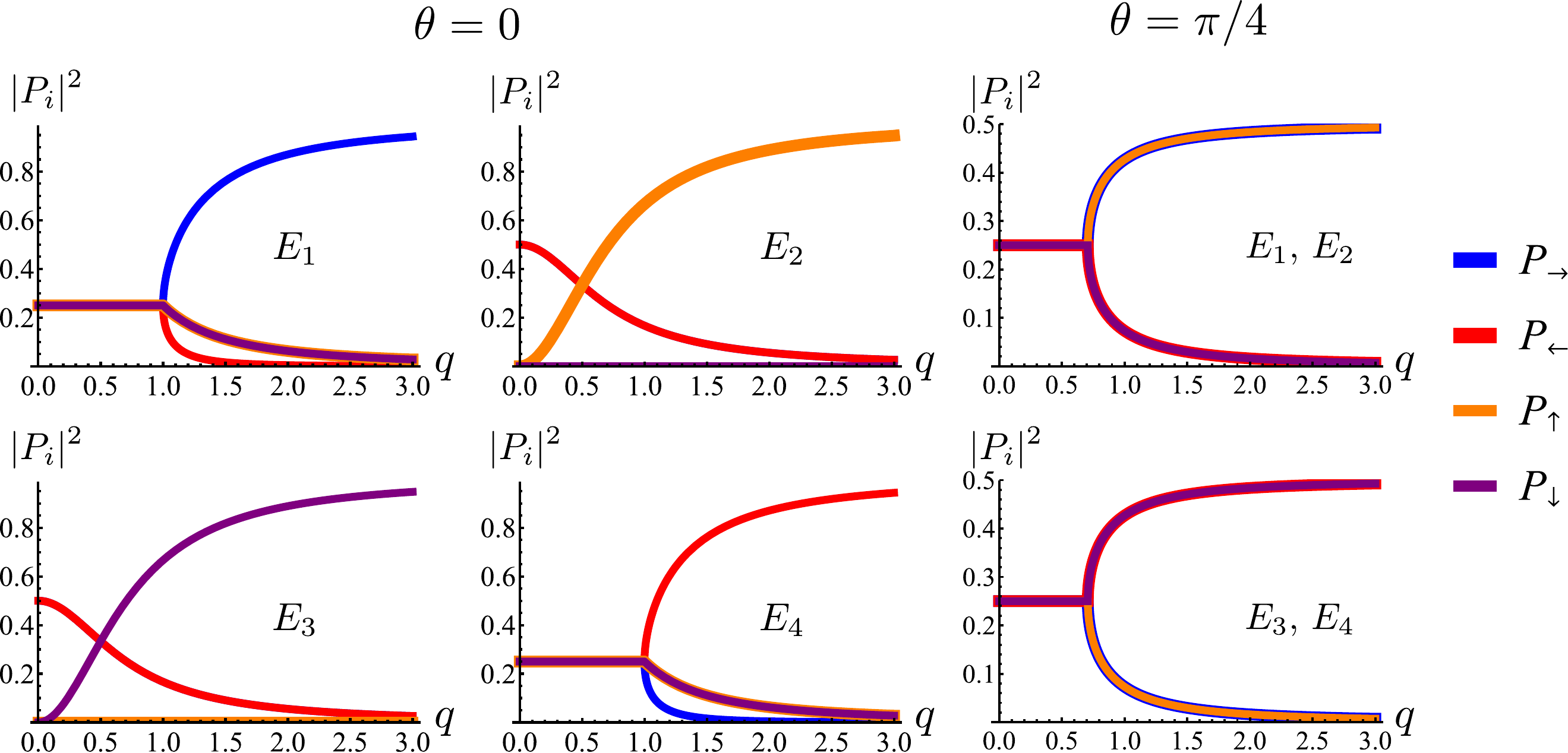}
    \caption{Square of the absolute value of the components of the eigenmodes of the  dynamical matrix in Eq.~(\ref{Eq:dyn_matrix}) as functions of the radial wave-number $q$ and its polar angle $\theta$
    \label{fig:eigenvectors}}
\end{figure*}
Much of the behavior of Eq.~(\ref{Eq:master_eq}) can be inferred from analyzing the spectrum of $\boldsymbol{H}_D$. Via Fourier transformation, the eigenvalues of  $\boldsymbol{H}_D$ are
 %\begin{equation}
%\label{Eq:eigenvalues}
%E(\boldsymbol{q}) = 1\pm \sqrt{\frac{1- q^2}{2} \pm \sqrt{\frac{1-2\,q^2+q^4 %\cos^2(2\theta)}{4}}},
%\end{equation}
\begin{equation}
    \label{Eq:eigenvalues}
    E(\boldsymbol{q}) = 1\pm \frac{1}{\sqrt{2}}\sqrt{1- q^2 \pm \sqrt{1-2\,q^2+q^4 \cos^2(2\theta)}},
\end{equation}
in which $q=|\boldsymbol{q}|$ and $\theta = \atantwo(q_y,q_x)$. These are depicted in Fig.~\ref{fig:eigenvalues}. The components of the associated eigenvectors are shown in Fig.~\ref{fig:eigenvectors}.

First, the spectrum is explicitly anisotropic as it has an explicit dependence on $\theta$, \add{which is} a consequence of the system's underlying four-fold symmetry. Furthermore, at long wavelengths, the system is completely diffusive as the spectrum is purely real and $E_1$, which is the only hydrodynamical mode (the only one that survives at long time scales), goes to zero as $E_1 \sim O(q^2)$ for $q\ll1$. At long wavelengths, all components, and hence all propagating directions, have the same weight (Fig.~\ref{fig:eigenvectors}). Thus the dynamics become isotropic at long times. 

More interesting is what happens at short length scales --- along the $x$ axis ($\theta$=0), each mode can be identified with one direction of propagation, as they only have one non-zero component (Fig.~\ref{fig:eigenvectors}). Moreover, although all modes decay at the same rate, only two acquire an imaginary part and can therefore propagate (Fig.~\ref{fig:eigenvalues}). These are precisely the modes associated with movement along the horizontal axis: $E_1$ and $E_4$. Furthermore, the different sign in the imaginary part of these modes ensures that $E_1$, associated with $P_\rightarrow$ has modes traveling to the right; whereas $E_4$, associated with $P_\leftarrow$, has modes traveling to the left. Since the system has a four-fold symmetry, the same picture is evident along the vertical axis but with propagation in the upward or downward directions. 

Along the diagonal, $\theta = \pi/4$, the components do not fully decouple. Instead, pairs of components have the same weight, the first pair being $P_\rightarrow$ and $P_\uparrow$, and the second $P_\leftarrow$ and $P_\downarrow$ (Fig.~\ref{fig:eigenvectors}, right column). These pairs also share the same sign for the imaginary part of their spectrum (Fig.~\ref{fig:eigenvalues}, bottom right), allowing the modes to travel toward the first and third quadrants, respectively. Regarding how fast these modes travel, it is seen from the asymptotic lines in the bottom row of Fig.~\ref{fig:eigenvalues} that they do so at a speed of $v_0$, the natural unit of velocity, which is equally distributed along two components along the diagonal. Combining these results, for a well-localized initial condition, the system initially exhibits propulsive behavior as the distribution travels in the direction associated with each component, propelled by the traveling modes at short wavelengths.  However, as particles start to turn, the modes begin to decay, with short wavelengths decaying faster than long ones. In the limit $t\gg T$, this discrepancy in decay rates results in only one isotropic mode surviving, with purely diffusive dynamics.

Finally, comparing simulations of this continuum model, Eq.~(\ref{Eq:master_eq}) with the statistics derived from agent simulations (Eqs.~(\ref{eq:Langevin_r})-(\ref{eq:Langevin_Phi})), in particular, the $\text{MSD}$, fourth moment and $\text{NGP}$ (Fig.~\ref{fig:moments}) reveals that the continuum model captures the essential features as the curves collapse. This reinforces the idea that when looking under the scale of $\ell$ and $T$, the dynamics are independent of the underlying details of the microscopic model and thus well described by the simplified continuum description.

\section{The Diffusive Regime: Long-wavelength Theory}

We have demonstrated two properties of the long-time dynamics: {\it (i)} diffusivity (Fig.~\ref{fig:moments}(a)) and {\it (ii)} isotropy (Fig.~\ref{fig:moments}(b)). However, it is not necessarily true that both properties emerge simultaneously. For example, intermediate-time dynamics may exhibit anisotropic diffusion or may be better described as isotropic propulsion. This section develops a hydrodynamic theory of the model to study the anisotropy of the diffusive regime and the approach to isotropy. 

The eigenvalue analysis of the spinor dynamics (Eq.~(\ref{Eq:eigenvalues})) revealed the presence of a hydrodynamic mode, suggesting this model supports a simplified long wavelength theory. To obtain it, notice that the particle density is $\rho = P_\rightarrow + P_\leftarrow + P_\uparrow + P_\downarrow $, the current is $\boldsymbol{j} = (P_\rightarrow - P_\leftarrow)\hat{\boldsymbol{e}}_x + (P_\uparrow - P_\downarrow)\hat{\boldsymbol{e}}_y $, and the degree of nematic ordering is described by $\chi = (P_\rightarrow + P_\leftarrow) - (P_\uparrow + P_\downarrow)$, which is positive if it is more likely for the particle to propagate along the horizontal axis and negative if it is more likely to propagate along the vertical axis.  With these definitions, the spinor dynamics (Eq.~(\ref{Eq:master_eq})) become
\begin{align}
    \label{eq:rho}
    &\partial_t \rho + \boldsymbol{\nabla} \cdot \boldsymbol{j} = 0\\
    \label{eq:j}
    &\partial_t \boldsymbol{j} + \frac{1}{2}(\boldsymbol{\nabla} \rho+ \boldsymbol{\nabla}_p \chi) = -\boldsymbol{j} \\
    \label{eq:chi}
    &\partial_t \chi +  \boldsymbol{\nabla}_p \cdot  \boldsymbol{j} = -2\chi,
\end{align}
in which $\boldsymbol{\nabla}_p \equiv (\partial_x, -\partial_y)$. Since the system does not possess particle-particle interactions that may lead to polar or nematic order, and $\rho$ is the only conserved quantity, $\rho$ is the only possible hydrodynamical variable. As such, $\boldsymbol{j}$ and $\chi$ act as fast variables, and their time derivatives are negligible \cite{Bertin2009}. Recursively substituting Eqs.~(\ref{eq:j}) and (\ref{eq:chi}) into Eq.~(\ref{eq:rho}) produces the following modified diffusion equation
\begin{equation}
    \label{eq:long_wave}
    \partial_t\rho = \frac{1}{2}\nabla^2\rho+\frac{1}{2^3}\nabla_p^4\rho+\frac{1}{2^5}\nabla^2\nabla_p^4\rho,
\end{equation}
in which the differential operator $\nabla_p^2 \equiv \partial_x^2-\partial_y^2$ serves as the source of anisotropy in the model. Through $\nabla_p$, the anisotropy appears on gradients of fourth order or higher, thus ensuring that any anisotropy in the system will be short-lived.  Although one ideally looks to truncate derivatives in the smallest possible order, it is necessary to go to the sixth order in this case. While truncating at the second order will just yield the diffusion equation, truncating at the fourth-order will result in instabilities arising at short wavelengths. As such, the sixth-order derivative is necessary to stabilize the equation at short wavelengths. 

\begin{figure}
    \includegraphics[width=\columnwidth]{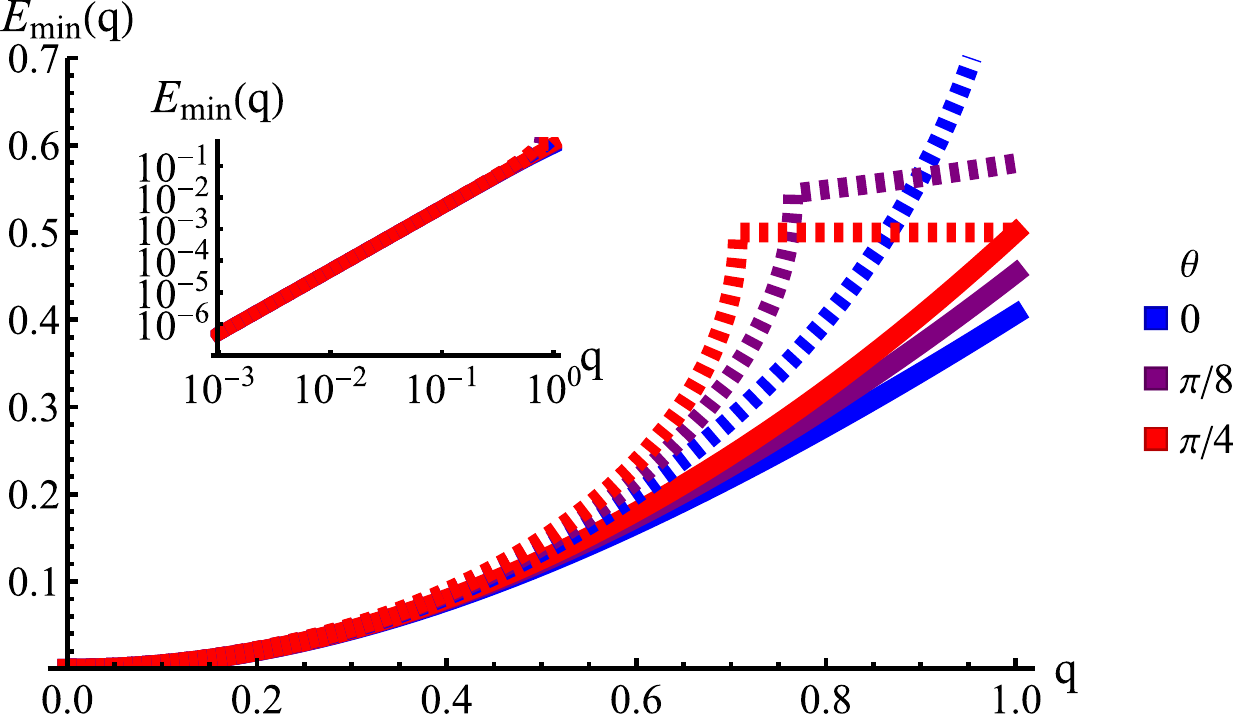}
    \caption{Lowest eigenvalue of Eq.~(\ref{Eq:eigenvalues}) (dashed lines) and Eq.~(\ref{eq:hydro_spec}) (solid lines) for different angles (blue: $\theta = 0$, purple: $\theta = \pi/8$ and red: $\theta = \pi/4$). Inset: Log-log scale, demonstrating that the curves collapse at small wave-numbers $q$, showing that the theory is accurate at small wave-numbers.
    \label{fig:comparison}}
\end{figure}

In Fourier space, the spectrum of Eq.~(\ref{eq:long_wave}) is given by
\begin{equation}
    \label{eq:hydro_spec_cart}
    \begin{split}
    E(q) = &\frac{1}{2}(q_x^2+q_y^2)-\frac{1}{2^3}(qx^4-2 q_x^2 q_y^2+q_y^4)\\
    &+\frac{1}{2^5}(q_x^6-q_x^4q_y^2-q_x^2q_y^4+q_y^6),
    \end{split}
\end{equation}
or, alternatively, in polar form ($q_x= q\cos(\theta)$,$q_y=q\sin(\theta)$)
\begin{equation}
    \label{eq:hydro_spec}
    \frac{1}{2}E(q) = \left(\frac{q}{2}\right)^2-\cos(2\theta)^2\left(\frac{q}{2}\right)^4+\cos(2\theta)^2\left(\frac{q}{2}\right)^6.
\end{equation}
Because of its direct dependence on $\theta$, $E(q)$ is explicitly anisotropic. 
To compare the reduced hydrodynamic theory with the more detailed continuum theory, we compare this spectrum with the lowest eigenvalue in Eq.~(\ref{Eq:eigenvalues}) (Fig.~\ref{fig:comparison}).
The hydrodynamic theory matches the lowest eigenmodes but loses the details of the behavior at short-length scales. Consequently, the hydrodynamic theory is only valid when describing phenomenology occurring at length scales larger than $1$, i.e., $q\ll 1$ (using $\ell$ as the unit of length). Furthermore, the theory is essentially diffusive as it loses all traveling modes. The fact that the model still presents anisotropic features while being fully diffusive reveals that the change to diffusive behavior precedes the change to the isotropic one.  
 
Having derived the theory, it can now be used to study how long the anisotropic features last by describing the time evolution of an initial probability distribution $\rho_0 = \rho(t=0)$
\begin{equation}
    \rho_0(\boldsymbol{r}) = \frac{1}{2\pi} \int d\boldsymbol{q} \hat\rho_0(\boldsymbol{q})e^{-i\boldsymbol{q}\cdot\boldsymbol{r}}.
\end{equation}
Because the theory is only valid at large length scales, we demand this initial distribution to have vanishingly small Fourier components at large wave-numbers, i.e., $\hat{f}(\boldsymbol{q})\approx 0$ for $q\gg 1$. For the sake of simplicity, we take $\rho_0$ to be a normal distribution of variance $\sigma^2\geq1$. For $\rho_0(\boldsymbol{r}) = \exp(-r^2/(2\sigma^2))/(2\pi s^2)$, we have $\hat{\rho} (\boldsymbol{q}) = \exp(-\sigma^2 q^2/2)/(2\pi)$, which is exponentially small at large $q$'s.  With this, the time evolution of the distribution is given by
\begin{equation}
    \label{eq:time_rho}
    \rho(\boldsymbol{r},t) = \frac{1}{2\pi}\int d\boldsymbol{q} \hat{\rho}_0(\boldsymbol{q})\, e^{-E(\boldsymbol{q}) t-i\boldsymbol{q}\cdot\boldsymbol{r}}.
\end{equation}
Because this integral is heavily suppressed for $q>1$, the limit $q\ll1$ is always satisfied in the integrand.  With this in mind, $q^2 \gg q^4, q^6$, which allows the approximation
\begin{equation}
    \begin{split}
        e^{-E(\boldsymbol{q})t} \approx  e^{-q^2 t/2}&\left[ 1+\left(\frac{1}{2^3}(q_x^2-q_y^2)^2\right.\right.\\
        &\left.\left.-\frac{1}{2^5}(q_x^6-q_x^4q_y^2-q_x^2q_y^4+q_y^6)\right)t\right]. 
    \end{split}
\end{equation}
In turn, the integral over $\boldsymbol{q}$ in Eq.~(\ref{eq:time_rho}) can be performed, yielding the time-dependent distribution
% \begin{widetext}
%     \begin{equation}
%         \label{eq:full_time_evolve}
%         \rho(\boldsymbol{r},t) = \frac{e^{-r^2/(\sigma^2+t)}}{2\pi(\sigma^2+t)}\left[1+\frac{t}{2(\sigma^2+t)^2}\left(1-\frac{(3+2 r^2)}{2(\sigma^2+t)}+\frac{r^2(9+r^2\cos^2(2\theta))}{4(\sigma^2+t)^2}-\frac{r^4(9+5\cos(4\theta))}{16(\sigma^2+t)^3}+\frac{r^6\cos^2(2\theta)}{16 (\sigma^2+t)^4}\right)\right], 
%     \end{equation}
% \end{widetext}
\begin{equation}
    \label{eq:full_time_evolve}
    \begin{split}
    \rho(\boldsymbol{r},t) = \frac{e^{-r^2/\gamma}}{2\pi\gamma}&\left[1+\frac{t}{2\gamma^2}\left(1-\frac{(3+2 r^2)}{2\gamma}+\frac{r^2(9+r^2c_2^2)}{4\gamma^2}\right.\right.\\
    &\left.\left.-\frac{r^4(9+5c_4)}{16\gamma^3}+\frac{r^6c_2^2}{16 \gamma^4}\right)\right], 
    \end{split}
\end{equation}
in which we let $\gamma\equiv\sigma^2+t$ and $c_n=\cos\left(n\theta\right)$ for brevity.
As a first estimate of how anisotropic this distribution is, notice that for large $t$, the leading anisotropic term in Eq.~(\ref{eq:full_time_evolve}) decays in time as $\sim \cos^2(2\theta)/t^3$, which appears to suggest that any anisotropic feature would be short-lived.  However, this point of view is local and does not tell the entire story. 

To study the distribution's global properties, consider its moments.
The second moment represents the mean squared displacement
\begin{equation}
    \left\langle r^2 \right\rangle = 2(\sigma^2+t).
\end{equation}
which confirms the diffusivity of the model. Besides the $\sigma^2$ term that originates on the variance of the initial distribution, we have that $\langle r^2 \rangle\sim t$ at all times. Surprisingly the diffusion constant is $D=1/2$, which is exactly the same result obtained by considering only the lowest gradients in Eq.~(\ref{eq:rho}). As such, the anisotropic features of the dynamics do not seem to introduce anisotropic diffusion. Indeed, the diffusion coefficient along any line on the plane is the same: $1/4$. This is because the local anisotropic dynamics do not make an explicit distinction between the Cartesian axes (see Appendix).  In summary, there are no anisotropic features at the level of the second moment.  

\begin{figure}
    \includegraphics[width=\columnwidth]{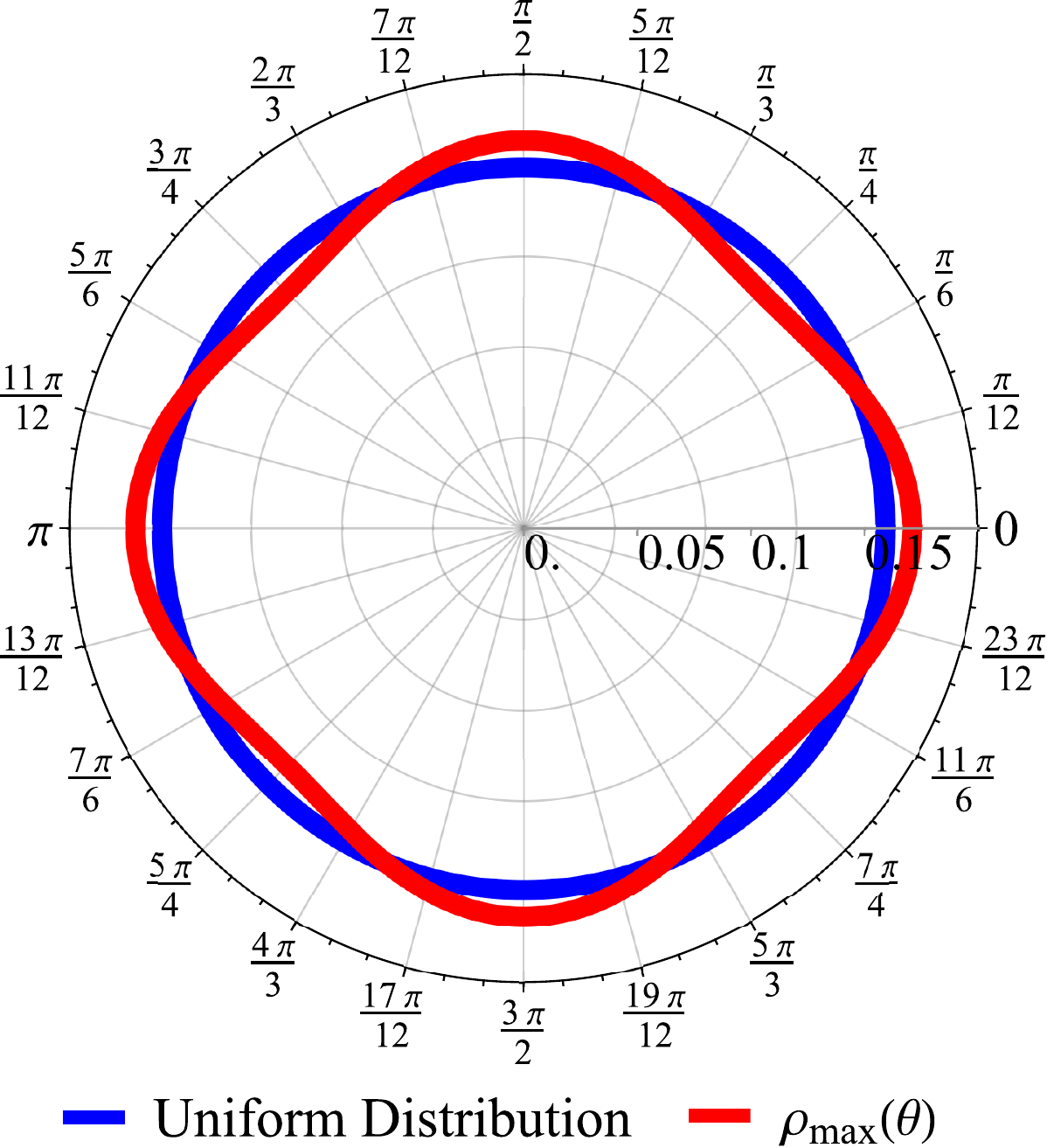}
    \caption{Angular distribution obtained from the long wavelength theory at the time of maximum anisotropicity $\rho_{max}(\theta)$ ($t=2$, $\sigma=1$). The distribution can be noticeably anisotropic even in the diffusive regime. The maximum anisotropicity is not achieved at $t=0$ (but rather $t=2$) because the initial distribution is isotropic.
    \label{fig:angular}}
\end{figure}

\begin{figure}
    \includegraphics[width=\columnwidth]{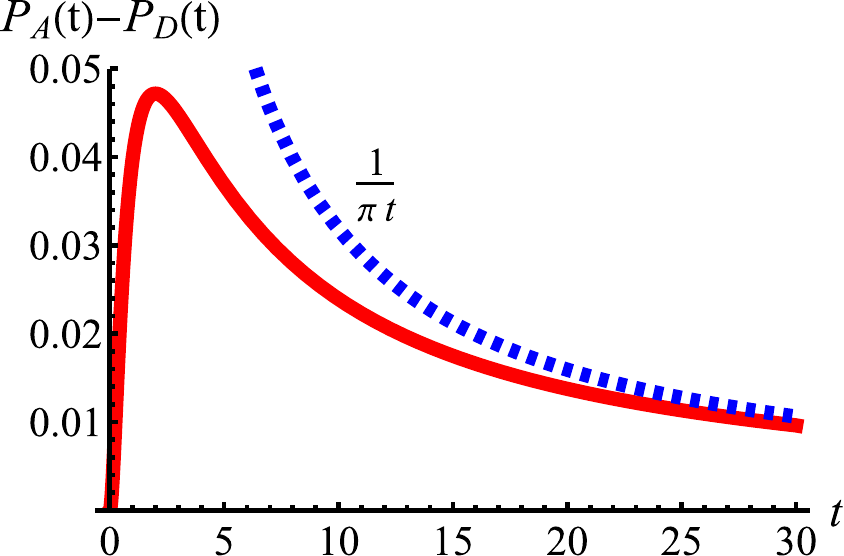}
    \caption{The difference between the probability of being closer to the axes $P_A$ and the probability of being closer to the diagonals $P_D$ as a function of dimensionless time $t$ (red curve). The blue dashed line is the long-time asymptotic behavior $\sim 1/(\pi t)$. The difference reaches its maximum at $t=2$ with an approximate value of $4.71\%$. It decays as $1/t$, keeping a difference above $1\%$ even after $t=30$.
    \label{fig:prob_diff}}
\end{figure}

However, this is not the case for the fourth moment
\begin{equation}
    \left\langle r^4 \right\rangle = 8 (\sigma^2+t)^2+4t,
\end{equation}
which is a different result than the one obtained considering only the lowest gradients (i.e., $8(\sigma^2+t)^2$). This is due to the anisotropic features contributing to deviating the distribution from a Gaussian. The non-Gaussian parameter ($\text{NGP}$) is
\begin{equation}
    \label{eq:NGP}
    \text{NGP} = \frac{1}{2}\frac{\langle r^4\rangle}{\langle r^2\rangle}-1 = \frac{t}{2(\sigma^2+t)^2}.
\end{equation}
First, in this case $\text{NGP}>0$, whereas for the master equation and agent simulations $\text{NGP}<0$ (Fig.~\ref{fig:moments}(c)). This is because, unlike the agent's simulations, which are driven by propulsive forces, the anisotropy, in this case, is carried out diffusively. Second, $\text{NGP}$ decays as $t^{-1}$, indicating that the anisotropic features actually persist much longer than initially suggested by the $t^{-3}$ decay observed in the distribution. This is corroborated by directly inspecting the angular distribution obtained by integrating the spatial distribution over the radial distance $r$  $\rho(\theta,t) = \int_0^{\infty} dr\, r \rho(\boldsymbol{r},t)$, which gives
\begin{equation}
    \label{eq:angular_dist}
    \rho(\theta,t) = \frac{1}{2\pi} + \frac{((\sigma^2+t)-1)t}{4\pi(\sigma^2+t)^3} \cos(4\theta).
\end{equation}
The non-monotonicity of $\rho(\theta,t)$ is a direct measure of the anisotropicity of the distribution (Fig.~\ref{fig:angular}). 
This non-monotonicity is provided by the second term in Eq.~(\ref{eq:angular_dist}). Interestingly, this term decays in time as $\sim t^{-1}$, just as the $\text{NGP}$, thus corroborating that this is the true rate of decay of the anisotropic features of the system, instead of $t^{-3}$.
Finally, the anisotropicity achieved by the system over time can be quantified by computing the probability of a particle being closer to the  axes rather than the diagonal $P_A$. Namely, four times the integral of $\rho(\theta,t)$ in the interval $[-\pi/8,\pi/8]$. Its reciprocal, $P_D$ is the probability of being closer to the diagonals, $P_D$. These are given by:
\begin{align}
    P_A = \frac{1}{2}+\frac{((\sigma^2+t)-1)t}{2(\sigma^2+t)^3}\\
    P_D = \frac{1}{2}-\frac{((\sigma^2+t)-1)t}{2(\sigma^2+t)^3}.
\end{align}
These show that $P_A>P_D$ (since $\sigma^2\geq 1)$, reflecting that the particles prefer to travel parallel to the axes. The difference between the two is depicted in Fig.~\ref{fig:prob_diff} for $\sigma=1$. This demonstrates that there is a temporal window on anisotropic dynamics. There is no difference at the beginning if the initial distribution is isotropic. However, anisotropy increases until it reaches a maximum of almost 5\% at $t=2$ (in units of $T$) driven by the anisotropic dynamics in the model. Anisotropy then decays as $1/t$, keeping at least a 1\% difference even after $t=30$.

\section{Conclusions}

Inspired by self-motile cells in anisotropic environments and developments of colloidal inclusions in active nematics, we have studied the dynamics of a particle following an ``run-and-tumble-turn" process, in which the particle is aware of its net alignment concerning the inherent frame of the environment and prefers to self-propel parallel to one of the cartesian axes. Through agent-based simulations of particles following a Langevin equation, the effects of noise and the width of the angular potential are explored at short and long time scales. The short-time effect is to renormalize the self-propulsion speed but keep the direction along the current axis. This process cannot induce a reversal in the direction of motion and thus only modifies the propulsive regime of the dynamics.  At longer time scales, the noise can allow the particle's orientation to overcome the angular potential barrier, inducing a right-angle turn.  This process governs the long-time dynamics of the system and determines the transition time from propulsive to diffusive behavior. 

We further explored how the angular potential's width affects the tumble-turn rate, the angular distribution's spread, and, therefore, how well collimated a beam of trajectories is. In particular, we learned that the width of the potential only alters the trajectories' squarishness at short time (length) scales. When looking at long-time (length) scales, all trajectories are indistinguishable for the range of parameters we considered. With this in mind, for the sake of simplicity, we considered the limit of a very narrow potential that keeps particles traveling mostly parallel to the axes. In this limit, we developed a continuum model based on a four-component spinor that details the probability of having a particle traveling along one of the cartesian axes. Solving this model numerically and comparing it with agent simulations demonstrates a good qualitative match with the agent-based simulations at long time scales, even for wider potentials, as the change to diffusive behavior seems to be independent of the details that drive it. 
Furthermore, the spectrum and eigenvectors of this continuum model characterize the role of long and short time scales over time and how the dynamics shift from propulsive to diffusive, and also how the latter behavior drives the system to a global isotropic behavior, despise its local anisotropicity. 
To study this transition further, we derived the model's hydrodynamic limit and solved it explicitly.  Although exhibiting a purely diffusive regime, the model can drive and sustain global anisotropic behavior, albeit for intermediate times. The long-time anisotropicity decays inversely with time. 

We hope our results will expand the current literature on active random processes, particularly in processes explicitly presenting four-fold symmetric tumbles \cite{Ruske2021, Loy2021, Loewe2022} or in which keeping anisotropic features on the global scale is important, such as biological processes occurring in anisotropic environments or that
exhibit inherent anisotropic features, as recent experiments in
morphology have shown how anisotropic division in shrimp embryos leads to a macroscopic four-fold symmetry \cite{Cislo2023}. We think our results can contribute to a better understanding of the mechanisms that help translate local anisotropies into macroscopic lengthscales and apply them to similar morphological processes. Finally, provided the recent interest in motility strategies in bacteria, we hope that by highlighting this route from local active anisotropies translate to larger scales, we can contribute towards the development of control protocols of living systems and the engineering of their transport processes.

%\subsection{Probabilistic description}

\section{Acknowledgments}

\begin{acknowledgments}
    This work was supported by the European Research Council (ERC) under the European Union’s Horizon 2020 research and innovation program (grant agreement no. 851196).
    For the purpose of open access, the authors have applied a Creative Commons Attribution (CC BY) license to any Author Accepted Manuscript version arising from this submission.
\end{acknowledgments}

\section{Appendix}

The anisotropicity of the long wave theory, Eq.~(\ref{eq:long_wave}) is subtle because it originated simply on the 4-fold symmetry of the underlying master equation, Eq.~(\ref{Eq:master_eq}). Since these dynamics do not break the symmetry between the four preferred directions (left, right, upwards, and downwards), the resulting hydrodynamic theory does not distinguish between the Cartesian axes, leading to isotropic diffusion in the lowest gradient  (as it can be seen in Eq.~(\ref{eq:long_wave}). Indeed, the anisotropic features arise at higher gradient order.  While in the main text, we focused on the consequences of this effect, in this appendix, we show how explicitly breaking the symmetry between the preferred directions causes the system to be explicitly anisotropic in that it develops a proper diffusion tensor. 

So far, we have considered that there is an equal probability rate of turning right or left than to turning upwards or downwards. This balance, however, is not a necessity of the dynamics. For example, if the four minima of the angular potential of our colloids have different heights, then by Krammer's formula, the particles would experience different tumble-turn rates. Such an effect, for example, can occur in the Janus colloid of reference \cite{Loewe2022}: the active force propelling the colloid modifies the potential that bounds it to its companion defect, which is responsible for its steering dynamics.  As such, the minima of the potential at the planar or homeotropic sides have different heights. Although this effect is negligible at small activities, it can be important as activity increases.  
Inspired by the above, we consider a particle whose lateral turning rate, $\lambda_L$, differs from its vertical one, $\lambda_V$. Taking $1/(\lambda_L+\lambda_V)$ as the unit of time and $v_0/(\lambda_L+\lambda_L)$ as the unit of length, this results in Eqs.~(\ref{eq:rho})-(\ref{eq:chi}) being modified to
\begin{align}
    \label{eq:rho2}
    &\partial_t \rho + \boldsymbol{\nabla} \cdot \boldsymbol{j} = 0\\
    \label{eq:j2}
    &\partial_t \boldsymbol{j} + \frac{1}{2}(\boldsymbol{\nabla} \rho+ \boldsymbol{\nabla}_p \chi) = -\boldsymbol{\Lambda} \boldsymbol{j} \\
    \label{eq:chi2}
    &\partial_t \chi +  \boldsymbol{\nabla}_p \cdot  \boldsymbol{j} = -2\chi,
\end{align}
in which we have defined 
\begin{equation}
    \boldsymbol{\Lambda} \equiv \frac{2}{\lambda_L+\lambda_V}  
    \begin{pmatrix}
        \lambda_L & 0 \\
        0 & \lambda_V 
    \end{pmatrix} 
    = \frac{2}{1+\alpha}
    \begin{pmatrix}
        1 & 0 \\
        0 & \alpha 
    \end{pmatrix},
\end{equation}
and $\alpha\equiv\lambda_V/\lambda_L$.
Proceeding as before, we obtain the following hydrodynamic theory
%\begin{equation}
%    \begin{split}
%        &\partial_t\rho = \frac{1}{4}\left(\frac{1}{\lambda_L}+\frac{1}{\lambda_V}\right)\left[\lambda_L\partial_x^2 + \lambda_V\partial_y^2\right]\rho\\
        %&+\frac{1}{32}\left(\frac{1}{\lambda_L}+\frac{1}{\lambda_V}\right)^2\left[\lambda_L\partial_x^2-\lambda_V\partial_y^2\right]^2\rho\\
        %&+\frac{1}{256}\left(\frac{1}{\lambda_L}+\frac{1}{\lambda_V}\right)^3\left[(\lambda_L^2\partial_x^2-\lambda_V^2\partial_y^2)(\lambda_L\partial_x^4-\lambda_V\partial_y^4)\right]\rho,
    %\end{split}
%\end{equation}
\begin{equation}
    \begin{split}
        \partial_t\rho &= \frac{1}{4}\left(1+\frac{1}{\alpha}\right)\left[\partial_x^2 + \alpha\partial_y^2\right]\rho\\
        &+\frac{1}{32}\left(1+\frac{1}{\alpha}\right)^2\left[\partial_x^2-\alpha\partial_y^2\right]^2\rho\\
        &+\frac{1}{256}\left(1+\frac{1}{\alpha}\right)^3\left[(\partial_x^2-\alpha^2\partial_y^2)(\partial_x^4-\alpha\partial_y^4)\right]\rho.
    \end{split}
\end{equation}
Setting $\alpha=1$ results in the same expression as Eq.~(\ref{eq:long_wave}). By rescaling lengths again by $\sqrt(1+1/\alpha)/2$, we finally get
\begin{equation}
    \label{eq:rescaled}
    \begin{split}
        \partial_t\rho &= \left[\partial_x^2 +\alpha\partial_y^2\right]\rho
        +\frac{1}{2}\left[\partial_x^2-\alpha\partial_y^2\right]^2\rho\\
        &+\frac{1}{4}\left[(\partial_x^2-\alpha^2\partial_y^2)(\partial_x^4-\alpha\partial_y^4)\right]\rho.
    \end{split}
\end{equation}
This equation explicitly breaks the symmetry between axes. Moreover, the anisotropy now enters at the leading order in gradients.  Indeed, we can readily read the diffusion tensor
\begin{equation}
    D = \begin{pmatrix}
    1 & 0 \\
    0 & \alpha 
    \end{pmatrix}.
\end{equation}
Beyond this anisotropic diffusion, Eq.~(\ref{eq:rescaled}) still presents higher order anisotropic terms originating in the rectangular-ness of the underlying dynamics.

\bibliography{Biblo}
\bibliographystyle{apsrev4-2}

\end{document}